# Energy Saving Strategy Based on Profiling

**Milan Yadav, Kanak Khanna**

The NorthCap University, Gurugram, Haryana, India

**ABSTRACT**

Constraints imposed by power consumption and the related costs are one of the key roadblocks to the design and development of next generation exascale systems. To mitigate these issues, strategies that reduce the power consumption of the processor are the need of the hour. Techniques such as Dynamic Voltage and Frequency Scaling (DVFS) exist which reduce the power consumption of a processor at runtime but they should be used in such a manner so that their overhead does not hamper application performance. In this paper, we propose an energy saving strategy which operates on timeslice basis to apply DVFS under a user defined performance constraint. Results show energy savings up to 7% when NAS benchmarks are tested on a laptop platform

**Keywords:** DVFS, Power Management, Frequency Scaling, Energy Saving

## I. INTRODUCTION

The ever increasing costs and constraints on power consumption are limiting the leap to the next generation exascale systems [34] as the power limit determined for these systems as per DoE guidelines is 20 MW. Therefore, there is an urgent need to limit the power consumption of modern computing systems to mitigate these issues.

To minimize the impact of the increasing power consumption on the computing performance, novel schemes/strategies need to be devised to reduce the power consumption of components within a computing system mainly processors without decreasing their performance by much. This means that a performance loss which can be tolerated for application execution can be pre-decided and then performance of the processor can be adjusted so to minimize power consumption. The current generation of Intel processors provides various P-states for dynamic voltage and frequency scaling (DVFS) [35][36] and T-states for introducing processor idle cycles (Throttling). For example, the Intel Haswell" micro-architecture provides fifteen P-states while the DRAM provides four frequencies. The delay of switching from one state to another depends on the relative ordering of the current and target states, as discussed, e.g., in [42]. The user may write a specific value to model-specific registers (MSRs) to change the P- or T-states of the processor. Intel micro-architecture starting from Sandy Bridge onwards estimates power and energy consumption of the CPU and memory through the built-in MSRs.

This paper proposes a power saving strategy which operates in a fixed timeslice fashion to reduce the energy footprint in modern processors during parallel application execution. The strategy makes use of DVFS in each timeslice to reduce processor frequency based on the performance constraint and workload. Results depict that up to 7% of energy can be saved.

The rest of the paper is organized as follows. Section 2 provides the background DVFS in Intel processors. Section 3 discusses the performance model that we use for the strategy. Section 4 provides the proposed energy saving strategy. Section 5 discusses the related works whereas Section 6 concludes the paper.

## II. Dynamic Voltage and Frequency Scaling

There are mainly two sources of power dissipation in digital CMOS circuits which are namely: Static Power [26] and Dynamic Power. Static Power is the power consumption which is consumed by a digital circuit when it is not in operation and it is mainly caused by leakage currents within the circuit. The other type of power consumption known as dynamic power consumption is mainly caused by the switching activity in the circuit when it is in operations. It is mainly dependent upon the operating frequency and voltage of the processor.

The dynamic voltage and frequency scaling (DVFS) mechanism reduces the operating frequency and voltage of the processor on-the-fly during application execution, thereby reducing the dynamic power consumption. DVFS is applied by writing a specific value to the IA32_PERF_CTL [1][2] model specific register (MSR) in Intel processors and it is an architectural register which means it is present in different generations of Intel processors with the same address.

## III. Applying DVFS

The execution time ($t_f$) of an application at a processor frequency f can be divided into two non-overlapping portions: On-chip time ($t_{on}$) and off-chip time ($t_{off}$) [41].

$$t_f = (t_{on})(f_{max})/f + (t_{off}) \quad \ldots(1)$$

Where $f_{max}$ is the maximum available processor frequency. The on-chip time scales linearly with the processor frequency accordingly whereas the off-chip time is not affected by the processor frequency.

This is due to the fact that off-chip time is predominantly memory accesses which runs in the order of many microseconds during which, the processor is simply waiting for the data and not doing much useful work. Therefore, this stall time can be exploited by simply reducing the processor frequency so that the power consumption of the processor can be reduced. This idea forms the basis of this work.

Basically, by monitoring the memory intensity, more specifically the memory accesses per instruction, we can assess the performance loss that would be caused on the application of DVFS on application performance. Therefore, if appropriate frequencies are chosen through DVFS, the performance loss can be minimized and substantial energy savings can be achieved. In the next section, we provide the details for the runtime strategy.

## IV. Energy Saving Runtime Strategy

| MAPI Range | Chosen Frequency |
|---|---|
| 0-0.004 <= | 2.4 GHz |
| >0.004-0.01 <= | 2.2 GHz |
| >0.01-0.04 <= | 1.6 GHz |
| >0.04 | 1.2 GHz |

Table I: MAPI ranges and the respective processor frequencies selected for the strategy.

For the strategy, we operate it on a timeslice basis such that a frequency is selected for the next timeslice in the previous timeslice based on the memory access per-instruction metric (MAPI) which is measured through the Intel processor performance counters.

Table I shows the range of MAPI values for a timeslice and the appropriate value of frequency that will be selected in case the actual MAPI value falls in that range. These values were obtained through extensive profiling on our hardware platform on which we executed several single and multi-threaded applications to notice the change in performance with varying processor frequencies. The performance degradation was noted down with the corresponding frequency and the MAPI value. A pattern was noticed for the range of MAPI values such that when the actual MAPI value was in that range, the performance loss for a particular frequency peaked at a particular value. This methodology was used to choose suitable frequencies for MAPI ranges as shown in Table I.

So, the strategy works as follows. The application can be divided into equally sized "k" timeslices. At the end of each timeslice, we measure the MAPI during that timeslice using the performance counters and record them in a register. To predict the MAPI for the next timeslice, we make use of a history-based predictor which averages "n" previous values of MAPI to predict its next value.

This predicted value is used in Table I to set the frequency accordingly for the next timeslice. In this manner, the strategy runs for the "k" timeslices, setting the frequency for each timeslice and reducing power and energy consumption. The algorithm I provides underlying pseudocode algorithm for the runtime energy-saving strategy.

Algorithm I

Parameters:
R => MAPI Register.
$M_{avg}$ => Average value of MAPI for past "x" timeslices.
1. Initialize MAPI Register R.
2. After "x" timeslices have passed, calculate $M_{avg}$.
3. Based on Table I, select the appropriate frequency based on the value of $M_{avg}$.
4. Go to Step 2 till "k" timeslices are finished.

## V. Experimental Results

We conducted experiments on a Desktop platform with Intel Core 2 Quad 6600, quad-core processor which has processor frequency ranging from 2.4-1.2 GHz. NAS NPB parallel benchmarks [4] were used for evaluation of the strategy. The power and energy consumption of the platform were measured by using the P3 4400 Kill-a-watt power meter and "*userspace*" governor was used throughout the experiments so that frequency decided by the *userspace* could be used.

**Table 1:** NAS NPB Benchmark Information

| CG (Conjugate Gradient) | Estimates the smallest eigenvalue of a large sparse symmetric positive-definite matrix using the inverse iteration with the conjugate gradient method as a subroutine for solving systems of linear equations. |

| FT (Fast Fourier Transform) | Solves a three-dimensional partial differential equation. |
|---|---|
| MG (Multigrid) | Approximates the solution to three dimensional discrete poisson equation using the multigrid method. |
| SP (Scalar Pentadiagonal) | Solves a nonlinear system of partial differential equations. |

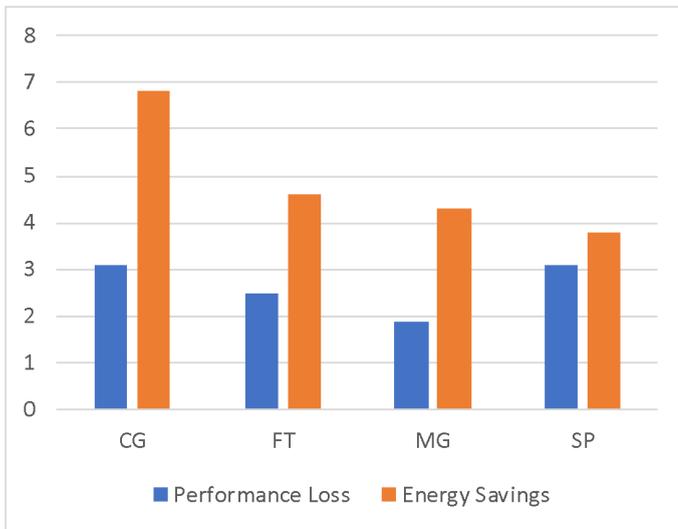

**Figure 1.** Performance Loss and Energy Savings for NAS NPB Benchmarks.

Four NAS NPB benchmarks namely CG, FT, MG and SP were chosen for the evaluation of the strategy which are explained in Table 1. Figure 1 shows the resultant performance loss and energy savings for the four chosen NAS NPB benchmarks. It can be observed from Figure 1 that for all the four benchmarks, the energy saving strategy does not result in a performance loss greater than 3%. Also, the average performance loss across the four benchmarks was determined to be ~2.4%.

In terms of energy savings, the largest amount of energy savings were obtained for the CG benchmark. This is due to the fact that CG benchmark was determined to be the most memory intensive out of the bunch with the MAPI figure staying above the 0.01 value. Therefore, it executes at frequencies of 2.2 and 1.6 GHz for different intervals. The other three benchmarks were not as memory intensive as CG, and depicted variable MAPI behavior through different timeslices, therefore, relatively lesser energy savings were obtained for them. Overall, the average energy savings for the four NAS benchmarks were ~5%.

## VI. Related Work

There are different strategies for reducing energy consumption in modern computing systems through DVFS. The first uses a fixed size timeslice based profiling methods with workload classification through performance counters [8], [10], [11], [14], [17], [19], [23],[24],[25],[30],[33],[37],[38],[39],[40],[41]. The other type identify communication phases which could be present in message passing etc. based communication intervals, to apply frequency scaling [7], [13], [15], [18], [20], [22], [25], [26], [27],[28],[29],[31],[32],[33]. While DVFS has been quite widely used to reduce the power consumption, it doesn't exactly provide the information regarding the instantaneous power consumption of the processor. Therefore, power limiting comes into picture so that power consumption of the processor can be directly controlled. Intel Running Average Power Limit (RAPL) [6] provides power clamping and energy metering capabilities starting from Intel SandyBridge processor generation. Runtime system "conductor" does power budgeting based on available power to different compute nodes and the communication slack available. Power limiting research focusing on capabilities of processor and DRAM along with a profiling based power budgeting strategy was proposed in [44]. Authors in [45] study and propose predictive models which facilitate forecasting values of performance parameters and the appropriate power limits for different components within a compute node to maximize performance.

## VII. Conclusions

The desire for achieving exascale performance has pushed the modern computing systems to operate at their maximum operating frequency and bandwidths. Consequently, the energy consumption and failure rates are also reaching prohibitive levels for these computing systems. To mitigate this issue, an energy saving strategy is proposed in this work which makes use of the DVFS capability present in the Intel processors to reduce the processor power and energy consumption. The strategy makes use of offline profiling to determine the frequency levels for different memory intensities and uses this information to scale the frequency for applications accordingly. Results on an Intel quad-core machine with four chosen NAS NPB benchmarks depict that energy savings of up to 7% can be achieved by using our strategy. Future work will explore a runtime performance model which would disable use of offline profiling and would enable the use to determine the frequency levels for an application at runtime.